# AN ESTIMATE OF THE GAS INFLOW RATE ALONG THE BAR IN NGC 7479

*A.C. Quillen,*[1,2] *Jay. A. Frogel,*[1,3] *Jeffrey D. P. Kenney,*[4] *R. W. Pogge,*[1]
*D. L. DePoy*[1]

September 13, 1994

Accepted for Publication in The Astrophysical Journal

[1] Ohio State University, Department of Astronomy, 174 W. 18th Ave., Columbus, OH 43210

[2] E-mail: quillen@payne.mps.ohio-state.edu

[3] Visiting Associate of the Observatories, Carnegie Institution of Washington

[4] Yale University, Department of Astronomy, P.O. Box 208101, New Haven, CT 06520-8101

## ABSTRACT

We present images of the barred galaxy NGC 7479 in the optical and near infrared broad–bands $B, V, R, J, H, K$, and images in H$\alpha$+[N II] and CO(J=1$\rightarrow$0) emission. The H$\alpha$ and CO emission in the bar are coincident and confined to narrow linear features that are offset from the center of the bar as observed in the near infrared. We estimate the gravitational potential from the $K$ image, which provides an estimate of the torque on the gas at the position of the CO emission in the bar. We find that the implied gas inflow velocity derived from the torque is $10-20$ km s$^{-1}$. Our inflow velocity is independent of the large streaming motions which can be observed in CO and HI.

## 1. INTRODUCTION

Numerical simulations show that non-axisymmetric distortions in the potential of a galaxy caused by galaxy interactions can be efficient at funneling gas into the nucleus of the galaxy (Mihos & Hernquist 1994a,b, Barnes & Hernquist 1991, Noguchi 1987, 1988, Hernquist 1989, Shlosman & Noguchi 1993, Shlosman, Begelman, & Frank 1990). During the merger, the galaxy exhibits an asymmetric spiral arm pattern, and has gas which can be undergoing star formation located in linear features that run next to a stellar bar. Because the linear features are offset from bar, there is a torque on the gas causing a loss in angular momentum and inflow towards the nucleus (van Albada & Roberts 1981).

NGC 7479, a barred galaxy with an asymmetrical two arm spiral, has many symptoms of interaction. The galaxy has an unusually large and well defined bar for an Sc galaxy (Blackman 1983, Elmegreen & Elmegreen 1985) and its nucleus appears to be offset from the center of mass of the system. It has a luminous nuclear region which is classified to be a LINER (Keel 1983), and as type HII by Fillipenko & Sargent (1985). The nucleus is compact at 10$\mu$m (Telesco, Dressel & Wolstencroft 1993) with IRAS colors between those of Seyfert and starburst galaxies (Roche et al. 1991). Beckman & Cepa (1990) observed dust lanes that "lead" the bar and "lag" the spiral arm pattern outside the bar; features predicted by Burbidge, Burbidge & Prendergast (1960) based on velocities observed along the bar. However, NGC 7479 is relatively isolated (Tully 1989), with nearest neighbor $\sim 2°$ away ($\sim$1 Mpc projected), so it is unlikely that a nearby galaxy has caused its present, short-lived asymmetric appearance. The appearance of the galaxy would however be consistent with a recent merger with a low mass companion (see Mihos & Hernquist 1994a).

In this paper we report the results of a multiband study of NGC 7479. This is a study of a strongly barred galaxy similar to the recent multiband study of the spiral galaxy M51 (Rix & Riecke 1993, Rand & Tilanus 1990). In §2 we present images of NGC 7479 at $B(0.44\mu$m), $V(0.55\mu$m), $R(0.71\mu$m), $J(1.25\mu$m), $H(1.65\mu$m), and $K(2.2\mu$m), as well as a narrow band H$\alpha$ +[NII] image and a high resolution CO(1-0) intensity map. As discussed in §3, we observe a thin linear feature in the broad band infrared color maps coincident with CO and H$\alpha$ emission that is offset from the center of the bar observed in the near infrared. Our data are a preliminary part of a survey of 200 to 300 galaxies that will produce a library of photometrically calibrated images of late-type galaxies from 0.4 to 2.2$\mu$m.

The numerical simulations of Mihos & Hernquist (1994a) predict that the asymmetric spiral arms and gas inflow should last only a few dynamical times. Because of the transient appearance of the NGC 7479 it is an interesting candidate in which to estimate a gas inflow rate. Gas inflow rates in barred galaxies are difficult to measure from velocities observed in CO and HI because of the large streaming velocities predicted in simulations (e.g. Athanassoula 1992) and also observed (e.g. Turner, Hurt & Hudson 1993). Typically the inflow velocities are predicted to be only a small fraction ($\lesssim 0.1$) of the radial streaming velocities (Athanassoula 1992). In §4 we estimate an inflow rate by computing the torque on the dense gas which depends upon its location in the gravitational field of the galaxy. The torque is derived from the gravitational potential which we estimate from our $K$ image following the procedure introduced in Quillen, Frogel & Gonzàlez (1994). This technique is well suited to our study since it estimates the gravitational potential of a non-axisymmetric system on the principal plane of the galaxy. We assume a distance to NGC 7479 of $D = 32$ Mpc which corresponds to a Hubble constant of 75 Mpc$^{-1}$ km s$^{-1}$. At this distance one arcsec corresponds to 160 pc.

## 2. OBSERVATIONS

### 2.1 *Near Infrared Images $J$, $H$, & $K$*

The $J$ and $K$ images were observed under photometric conditions on 1992 Nov. 15 with the 1.8m Perkins Telescope of the Ohio State and Ohio Wesleyan Universities at Lowell Observatory in Flagstaff, AZ using a $256 \times 256$ HgCdTe array in the Ohio State Infra-Red Imaging System (OSIRIS) (Depoy et al. 1993). The array covers a field of $6.6 \times 6.6$ arcminutes, with a spatial scale of 1.50 arcsec/pixel. Individual images were taken with an exposure time of $\sim 3.15$ seconds. The total on source integration time was 25.2 minutes in both bands. On 1992 Nov 16 an additional observation at $K$ was obtained with a total integration time of 20 minutes under non-photometric conditions.

An $H$ image and short exposures in $J$ and $K$ were obtained under photometric conditions on the 1.0m Swope Telescope at Las Campanas Observatory on 1993 Oct. 3. The infrared array (Persson et al. 1992) covered a field of $3.9 \times 3.9$ arcminutes, with a spatial scale of 0.92 arcsec/pixel. Individual images were taken with an exposure time of 30 seconds in $J$ and $H$, and 20 seconds in K. Total on source integration times were approximately 21 minutes in H, 3.5 minutes in J and 100 seconds in $K$. The short exposures at $J$ and $K$ were used to check the calibration of the longer exposures from the Perkins Telescope.

For both sets of observations, the sky was observed for a total integration of about half of the total on source integration time. Flat fields were constructed from median filtered sky frames. Images were aligned to the nearest pixel and combined after a slight non-linearity correction, flatfielding and sky subtraction. A planar surface was removed from the resulting images to correct for problems in sky subtraction probably due to scattered light.

The final images were calibrated using standard stars from Elias et al. (1982) to convert to the CTIO/CIT system. We compared the resulting images to aperture photometry by Willner et al. (1985) with an aperture diameter of $5''$ and by Devereux (1989) with aperture diameters of $5''.3$ and $9''.3$. We find that our $J$, $H$ and $K$ magnitudes are lower than those of Devereux (1989) by 0.12 mag and higher than that of Willner et al. (1985) by 0.44 mag. We agree with colors observed by both sets of authors. In contrast, the absolute calibration of our Las Campanas and Perkins data agreed to within $\pm 0.01$ mag at $J$ and $\pm 0.04$ mag at $K$.

In our images the noise per pixel has a variance that corresponds to 23.1 mag per arcsec$^2$ at $J$, 21.7 mag per arcsec$^2$ at $H$, and 22.1 mag per arcsec$^2$ at $K$. We note that at a given isophote level the $J$ and $K$ have approximately the same signal to noise whereas the $H$ image has a much lower signal to noise. The FWHM of stars in the images are $\sim 2''.6$ and are spatially undersampled. Near infrared images are displayed in Figures 1 and 2.

### 2.2 *Optical Images $B$, $V$, & $R$*

The $B$, $V$, and $R$ images were observed under photometric conditions on UT 1993 Sept 17 using the Ohio State University Imaging Fabry-Perot Spectrometer (IFPS) on the 1.8m Perkins Telescope. We used the IFPS in its direct imaging mode (no etalon in the beam) with the Lowell NSF Texas Instruments $800 \times 800$ CCD detector. With this detector, the focal reducing optics of the IFPS gives an image scale of $0.49''$/pixel with a slightly vignetted $6.5'$ field of view. Individual images were taken with an exposure time of 60 seconds in R , 120 seconds in V and 300 seconds in $B$. Total integration times were 5 minutes at $R$, 6 minutes at $V$ and 10 minutes at $B$. Flat fields were constructed from means of images taken at twilight. Images were combined after flatfielding, dark subtraction and a linear overscan bias subtraction. Cosmic rays events were removed using interactive median filtering. Noise in the image corresponds to $\sim 25.5$ mag/arcsec$^2$ in the three optical bands.

Calibration was done by observing stars in 2 Landolt fields (Landolt 1992). We compared magnitudes measured in synthetic round apertures centered on the nucleus in our $B$ and $V$ images to those listed listed in Longo & deVaucouleurs (1983). We find that our $B$ and $V$ zero points agree within 0.05 mag to a best fit to all magnitudes listed in Longo & deVaucouleurs (1983). Optical images are displayed in Figures 1 and 2.

### 2.3 *Color maps*

Registration for the color maps was done by centroiding on stars in the field. We used linear transformations which included a rotation and a scaling term. Images were transformed to the scale of the $1.50''$/pixel $J$ and $K$ images which have the largest pixels. We estimate that our registration is good to within 0.3 arcsec based upon the variance of centroids in different images of the field stars of the registered images. Prior to division the images were smoothed so that stars had the same FWHM as the J image. Color maps are displayed in Figures 3 and 4.

### 2.4 *Hα+[N II] Narrow Band Image*

The Hα+[N II] and continuum band CCD images were obtained on UT 1993 Sept 18 using the IFPS on the 1.8m Perkins Telescope. Two 50Å interference filters located below the $f/17$ cassegrain focus were tilt-tuned to the appropriate bandpasses for the redshifted Hα+[N II]λλ6548,6583 emission lines and a region of line-free continuum ∼69Å blueward of the emission-line band. Changes in the filter bandpasses due to placement in a slow non-parallel beam are insignificant (<0.01%). Conditions were non-photometric, with thin cirrus visible at sunset, and the seeing was steady at $\sim 2''$ FWHM measured from field stars on the images. Total exposure times were 1800 seconds (30 minutes) in each band. Flat field frames were constructed by combining twilight sky images. The images were flatfielded and debiased, and then registered, scaled, and subtracted to produce a pure emission-line image following the procedures described in Pogge (1992).

The flux was calibrated by summing over the entire image and comparing to Kennicutt & Kent (1983)'s integrated Hα+[N II] flux of $1.4 \times 10^{-12}$ ergs cm$^{-2}$ s$^{-1}$. The calibration of our image is uncertain by ±15% primarily due to uncertainties in the sky levels and the scaling factor of the continuum image. The residuals from stars in the field were not removed when summing the flux in the image but they contribute less than a few percent. The Hα image is displayed in Figures 3,4 and 5. The emission is confined to narrow features that follow the bar and spiral arms.

### 2.5 *OVRO Interferometer CO Observations*

Three fields covering the bar and nucleus of NGC 7479 were observed in the CO(J=1→0) line with the 3-element Owens Valley Radio Observatory millimeter-wave interferometer (Padin et al. 1991) between December 1987 and February 1989. The central field was centered at the Dressel & Condon (1976) position; the others were offset by $33''$ north and south.

All of the 10.4-meter telescopes (HPBW=$65''$) were equipped with cryogenic SIS receivers, whose noise temperatures were measured to be ∼300 K SSB. Measurements of an ambient temperature chopper wheel were used to correct for variations in sky opacity and receiver gain. The quasar 3C454.3 was used to calibrate the phase, and Uranus and Mars were used as absolute flux standards. Spectral coverage was provided by 32 5-MHz filterbank channels, resulting in a resolution of 13 km s$^{-1}$ and an instantaneous bandwidth of 416 km s$^{-1}$. Each field was observed in 4 different telescope configurations, with projected baselines ranging in length from 10 to 60 meters. After calibrating the data, channel maps were made using uniform weighting with the AIPS task MX. The channel maps from the 3 fields were combined with the AIPS mosaicing task MMAP.

In the final mosaic map, data from the central field was given half the weight of the other 2 fields and clipped beyond a radius of $20''$ (the 75% power point), since it was significantly noisier. Data from the other fields were clipped beyond a radius of $40''$ (the 35% power point). The resultant maps were created with a synthesized beam of $8.4'' \times 6.6''$ at PA= 0°, have sensitivity to structures as large as $30''$, and are corrected for the response of the primary beam. CO emission is detected in 26 of the 32 individual channel maps over the velocity range 2190 to 2502 km s$^{-1}$ (LSR) at a level $\geq 4\sigma$. The channels with emission were combined to form a moment 0 map of the total CO intensity. Figure 5a shows a CO contour map of a $30'' \times 115''$ region containing the nucleus and bar. Nothing from outside this region has been detected at a level $\geq 3\sigma$.

CO emission is detected from a marginally resolved circumnuclear component, and from several components along the bar. The OVRO map contains a CO flux of 150±20 Jy km s$^{-1}$ from the circumnuclear component (r=0-8"), and 120±30 Jy km $s^{-1}$ from the bar components (r=8-60"). From a comparison with a single dish CO map obtained with the FCRAO telescope (Kenney, unpublished), we estimate that the interferometer has detected 90±5% of the total circumnuclear flux, and 30±10% of the bar flux. The CO emission peak, which coincides with the CO kinematic center, is located at $\alpha(1950)=23^{\rm h}02^{\rm m}26.38^{\rm s}$, $\delta(1950)= 12^\circ\ 03'10.6''$, which is offset by $7''$ from the Dressel & Condon (1976) position.

## 3. RESULTS

### 3.1 *Morphology*

Near-infrared images of galaxies detect light primarily from cool giants and dwarfs that contribute a major fraction of the bolometric luminosity of a galaxy. Particularly in spiral galaxies, these stars are much better traces of the mass distribution of the galaxy than are the bluer, hotter stars (Aaronson 1977, Frogel 1988). Because extinction from dust is far less in the near-infrared than in the optical ($A_K \sim 0.1 A_V$) near-infrared images will also show more accurately the intrinsic shapes of galaxies. The asymmetric appearance of the spiral arms of NGC 7479 in the near infrared bands implies that the mass distribution of the galaxy is not centered about the nucleus. We note that the outermost isophotes (see Figure 2) are also not centered on the nucleus. The large size of the southern spiral arm, causes the the centroid at $K$ to be $\sim 5''$ south of the nucleus. This situation is typical of mergers (e.g. Mihos & Hernquist 1994a,b) and should last only a few rotation periods. We see no strong feature at $K$ that could have been the nucleus of a smaller galaxy so we suspect that the smaller galaxy has been disrupted.

NGC 7479 has spiral arms and bar that are the most sharply delineated in H$\alpha$ of the galaxies observed in the complete H$\alpha$ imaging survey (Pogge 1989) of 91 northern non-Seyfert spiral galaxies with magnitudes at $B$ greater than 12 magnitude. This galaxy also has the most highly asymmetric spiral arm pattern of any galaxy in this survey. The H$\alpha$ emission can be described in terms of two separate pairs of nearly linear features; a pair of linear features that run along the bar on either side of the nucleus, and an outer pair that extend into the spiral arms. A cartoon figure of the two pairs of linear features is shown at the lower right hand corner in Figure 4. The two pairs of linear features may be caused by two timescales of non-axisymmetric or quadrupolar distortions of the potential; an outer mode caused by the remnant of the smaller galaxy that has merged with NGC 7479 and produces the features seen as spiral arms, and an inner mode that we see as a bar.

In addition to the narrow feature seen in the $J-K$ color map, H$\alpha$ and CO(1-0) emission (see Fig. 3,4,5), there is a broader lens shaped region $\sim 100''$ long and $\sim 25''$ wide and with nearly constant red colors in all color maps (see Fig. 4) which is broader than the bar itself (compare Fig. 1 and 4). This region has sharp edges on the north-west and south-east side and is bounded by the outer pair of linear features of H$\alpha$ emission that extend into the spiral arms (see Figure 4). Colors outside the edge of the lens shaped region are blue (see Table 1 under the column out. ridge) which suggests that the edge of the region is not caused by a drop in extinction but caused by blue stars that are associated with the outer pair of linear shocks.

The bulge is small, and there is no radius at which the isophotes are round. Mihos & Hernquist (1994a,b) predict that in a minor merger, galaxies without bulges develop bars shortly after their first close passage. The asymmetric appearance, small bulge, and large bar in NGC 7479 fit this prediction. In order for there to be a significant bulge component in the region where the isophotes are not round, the bulge would have to be non-spherical and have an anisotropic velocity dispersion.

### 3.2 *The linear Shock-like Feature*

The bar of NGC 7479 contains a strikingly narrow linear feature that is easily visible in the $J-K$ and $H-K$ color maps (see Fig. 3), but difficult to detect in a $J-H$ color map. This feature runs parallel to but displaced from the major axis of the bar, in the sense that it is closer to the leading edge of the bar. Aside from the nucleus, the feature has the reddest $JHK$ colors of any part of the galaxy (see columns labeled lin. feature and nucleus in Table 1). It is also coincident with the location of the reddest colors observed in $R-J$ (see Fig. 3). As may be seen in Figure 4 this linear feature also contains spots with blue optical colors $B-V\sim 0.74$. H$\alpha$ and CO(1-0) emission are also strongly concentrated along this feature.

Although the S/N of our $H$ image is significantly lower than that of our $J$ and $K$ images, we can still estimate the average *difference* in color between the linear feature and the part of the bar immediately adjacent to it (see the column labeled lens in Table 1). The difference in $J-K$ (see Table 1) is about $0.1\pm 0.02$ and in $J-H$ it is about $0.03\pm 0.03$. It is unlikely that such a difference in color could arise from simple reddening since the ratio of color differences $\Delta(J-K)/\Delta(J-H)$ is typically $\sim 1.5$. We note that the apparent spatial coincidence of the H$\alpha$ and CO(1-0) emission and the blue spots (in $B-V$) points to a close association of young blue stars and cold dense molecular gas with the linear feature seen in the $J-K$ and $H-K$ color maps. There must be an additional emission process contributing to the $K$ surface brightness at a level that is $\sim 10\%$ above that of the bar, but with only a very small contribution at $J$. Possible emission processes that could contribute at $K$ are emission from giants and supergiants, dust emission (e.g. Sellgren 1984) and continuum emission from ionized gas. Studies of star forming regions have found that even in starbursts nebular emission contribute less than 1% to the $K$ band surface brightness (e.g. Lester et al. 1990 in M82), however emission from newly formed giants and supergiants can make a significant contribution to the $K$ surface brightness (Rieke et al. 1980, Lester et al. 1990, Krabbe et al. 1994). Higher resolution and better signal to noise near infrared images would resolve some of the ambiguities seen in our infrared color maps. This, coupled with spectral observations of hydrogen lines, a measurement of the CO index and far infrared images could determine the emission processes. Studies of M51 show that the H$\alpha$ emission is offset from the CO emission and regions of highest extinction (Rand & Kulkarni 1990). Thus it is quite possible that higher resolution images would show more structure and that different processes may dominate at different locations.

East-West slices across the bar at a number of vertical offsets are shown in Figure 6. As pointed out previously, we see that the H$\alpha$ and CO(1-0) emission are offset from the long axis of the bar defined in the near infrared (see Fig. 5). Within the bar corotation radius, the gas angular rotation rate is expected to be faster than the bar angular rotation rate. When a shock occurs, it is expected to occur on the "leading" side of the bar since in the frame in which the bar is stationary this side is downstream from the peak density of the bar. In the spiral arms of M51 (Rand & Kulkarni 1990) and at the bar end in M83 (Kenney & Lord 1991), H$\alpha$ emission occurs downstream from the CO(1-0) emission presumably because of a delay in the time between forming stars in the dense gas and creating HII regions. We do see this at the outer ends of the bar in NGC 7479 (see Fig. 5a). The larger lag in the outer region may be due to a faster passage of the gas through the shock at this radius, or because streamlines could be more circular near the end of the bar. Higher resolution CO and H$\alpha$ observations may show that they are also not coincident in the central region.

## 4. ESTIMATING THE GAS INFLOW RATE

### 4.1 *Estimating the Gravitational Potential*

We correct for the inclination of the galaxy assuming an inclination of $45° \pm 4$ and a position angle of $39° \pm 1$ (Grøsbol 1985). Following Quillen et al. (1994) we estimate the gravitational potential by convolving the deprojected $K$ image of the galaxy (with stars removed) with a function that depends on the vertical scale height of the disk. This technique is well suited to our study since it estimates the gravitational potential of an non-axisymmetric system in the principal plane of the galaxy and can be done quickly with a Fast Fourier Transform. We assumed a sech$^2$ form for the vertical density of the disk ($\rho_z(z) \propto \mathrm{sech}^2(z/h)$) where $h$ is the scale height of the disk. We used a vertical scale height $h = 4''$, that is $\sim 1/12$ times the exponential disk length of the galaxy $49.6'' \pm 2$ Grøsbol (1985). This assumption is consistent with optical and near infrared observations of edge-on disks (Barnaby & Thronson 1992, Wainscoat, Freeman & Hyland 1989) and resulted in a self consistent potential model for the bar of NGC 4314 (Quillen et al. 1994). Because of the small size of the bulge, we did not consider its three dimensional nature; thus our potential will not be accurate near the nucleus ($r \lesssim 8''$).

The gravitational potential in the plane of the galaxy, $\Phi(x, y)$, is shown in Figure 7 as a contour map of the quantity

$$\Phi(x,y)\left(\frac{M/L_K}{1.35 M_\odot/L_{K\odot}}\right)^{-1}\left(\frac{D}{32\mathrm{Mpc}}\right)^{-1} \tag{4.1}$$

in units of (km s$^{-1}$)$^2$ or (pc/Myr)$^2$. where $M/L_K$ is the $K$ band mass-to-light ratio with $L_K$ in units of solar $K$ luminosities, and $M$ in units of solar masses. Duval & Monnet (1985) observed in H$\alpha$ at a radius of $50''$ (where the rotation appears to be circular) a maximum line-of-sight velocity of 160 km s$^{-1}$, which corresponds to a circular velocity of 226 km s$^{-1}$ for a galaxy inclination of 45° (Grøsbol 1985). The $K$ band mass-to-light ratio we require in order to match this circular velocity is $M/L_K \approx 1.35 M_\odot/L_{K\odot}$. This corresponds to the mass-to-light ratio of a single burst population model of age 12 Gyr for metalicity [Fe/H]$= -0.25$ (Worthey 1994) which is a quite reasonable value for an average mass-to-light ratio. Our predicted rotation curve and angular rotation rate derived from the azimuthally averaged potential with this mass-to-light ratio is displayed in Figure 8. We find that at least one Inner Lindblad Resonance (ILR) exists, and that it must be very close to the nucleus. The location of the strong circumnuclear CO emission is near of inside the predicted location of the ILR as is observed in many other barred galaxies (Kenney et al. 1992). The small radius of the ILR and central concentration of molecular gas would be consistent with a merging scenario where the molecular gas in the nucleus now observed resulted from inflow. Assuming that corotation is near the bar end, we estimate that the bar angular rotation rate is $\Omega_b \sim 0.1$ Myr$^{-1}$ from the angular rotation rate at a radius of $\sim 45''$ near the end of the bar.

## 4.2 *Estimating the Inflow Rate*

Numerical simulations of gas flow in barred galaxies have found that the gas inflow primarily occurs in the dense gas associated with linear shocks which run along the bar (Athanassoula 1992, Hernquist 1989). Since the CO(1-0) is located in a linear feature we expect that the dense gas is in a shock and has a low non-radial velocity in the frame in which the bar is stationary (van Albada & Roberts 1981, Athanassoula 1992). We estimate the angular momentum per unit mass of the gas to be $l \approx r^2\Omega_b$ where $\Omega_b$ is the bar angular rotation rate. The torque per unit mass at the location of the molecular gas is $\tau = d\Phi/d\theta = dl/dt \approx (dr/dt)2r\Omega_b$. This results in an inflow speed of

$$\frac{dr}{dt} \approx \frac{d\Phi/d\theta}{2r\Omega_b}. \tag{4.2}$$

We estimate the inflow speed by summing the torque across the CO feature. The resulting inflow speed is displayed in Figure 9. A more accurate calculation of $dr/dt$ would need to consider the gas flow over the whole region. Since the $K$ image may have a component of non-stellar emission (see §3.2) we also computed $dr/dt$ using a potential generated from the $J$ image. The difference in the predicted inflow speeds was small, less than 1 km s$^{-1}$.

We estimate that the inflow speed lies in a range $10 - 20$ km s$^{-1}$ (see Fig. 9). The order of magnitude of the inflow speed measured here agrees with that predicted by the recent simulations of Mihos & Hernquist (1994) (C. Mihos private communication), but is higher than that predicted by Athanasoula (1992) ($1 - 6$ km s$^{-1}$). We note that we have summed the torque on the CO observed in the linear feature. It is possible that a sum over the whole gas distribution could result in a lower estimate for the inflow rate. The low resolution of the CO data causes the estimate to be inaccurate, particularly at small radii. To estimate the effect of this we estimated the inflow rate using the H$\alpha$ image (which has a smaller spatial resolution than the CO map) to trace gas mass, and found good agreement between the inflow speed predicted from the H$\alpha$ emission and that predicted from the CO (see Fig. 9). The H$\alpha$ emission is farther offset from the center of the bar for $r > 30''$ (see Fig. 5) and so the inflow rate predicted from the H$\alpha$ at this radius is larger than that predicted from the CO. If the vertical scale height ($h$) of the galaxy is smaller than we have assumed then the inflow speed should be higher. Quillen et al. (1994) found that the quadrupole component of the potential scaled approximately as $1/h$ so the inflow rate should scale approximately in the same way.

We estimate the mass inflow rate $dm/dt$ using the the empirical relation for the $H_2$ column density

$$\Sigma(H_2)(M_\odot \mathrm{pc}^{-2}) = 470 I_{\mathrm{CO}(1-0)}(\mathrm{Jy\ km\ s}^{-1}\mathrm{arcsec}^{-2}) \tag{4.3}$$

(Kenney et al. 1992) where $\Sigma(H_2)$ is the mass surface density of $H_2$ and $I_{\mathrm{CO}(1-0)}$ is the integrated intensity in the CO(1-0) line. This estimate is consistent with observations of optically thick molecular clouds in the Galactic disk (Bloemen et al. 1986). If we assume that the total $H_2$ + He mass is approximately 1.35 times the $H_2$ mass then for the gas column densities derived from the CO(1-0) emission we estimate a mass inflow rate of $4 \pm 2 \times 10^6 M_\odot$/Myr. We note that we have detected only 30±10% of the single dish flux in the bar (see §2.5). If the missing flux is uniformly distributed then both the inflow speed and mass inflow rate are still correct, however if the missing flux is in a broad component centered on the detected components, then we find approximately the same value for the inflow speed but could have three times as high a mass inflow rate.

Using the conversion factors listed above, we find that the total mass ($H_2$ + He) at the nucleus is $\sim 2.3 \times 10^9 M_\odot$. If we assume that all the gas in the nucleus resulted from an inflow at the rate as we have estimated, then our estimate implies a timescale since the inflow began of $\sim 6 \times 10^8$ years ago. The time for one rotation at the end of the bar $r \sim 45'' = 7.2$kpc (corotation) is $\sim 2.0 \times 10^8$ Myr so that our inflow timescale is approximately 3 rotations at the corotation radius. This timescale for inflow is consistent with the simulations of Mihos & Hernquist (1994a). We note that the timescale for a small galaxy to disrupt in a merger is typically a similar timescale, a few rotation periods (e.g. Hernquist 1989), depending upon the core radius of the small galaxy.

## 5. SUMMARY AND CONCLUSION

In this paper we have conducted a broad band study of the barred spiral galaxy NGC 7479 from 0.4 to 2.2$\mu$m. The galaxy has a long narrow region offset from the major axis of the bar (observed in the near infrared) which contains cold molecular gas detected in CO(1-0) emission and is also the site of star formation as is seen in H$\alpha$ emission. Colors in the near infrared are quite red suggesting that this region also contains emission at $K$. This region also contains some blue spots from associations of young stars, as well as regions of higher optical extinction. Outside the bar there is a separate set of linear features observed in H$\alpha$ and color maps that suggest there is a larger scale quadrupolar gravitational perturbation that has a slower pattern speed than the bar, possibly associated with the remains of a galaxy that has merged with NGC 7479.

We use our $K$ image to estimate the gravitational potential of the galaxy and then estimate an inflow speed based upon the torque on the dense gas as traced in CO emission. We estimate an inflow speed of $10 - 20$ km s$^{-1}$ and a mass inflow rate of $4 \pm 2 \times 10^6 M_\odot$/Myr. The inflow speed is consistent with the predictions of recent simulations of galaxy mergers (Mihos & Hernquist 1994a,b) and somewhat higher than the simulations of a steady state bar model (Athanassoula 1992). Our estimate is the first estimate (of which we are aware) of a bar mediated inflow rate that is based on observations. Because our estimate is derived from the torque on the gas, it is not affected by the large streaming motions that can be observed in CO and HI.

Higher resolution and better signal to noise near infrared images would resolve some of the ambiguities seen in our infrared color maps. This, coupled with spectral observations of hydrogen lines, a measurement of the CO index and far infrared images could determine the emission processes that could be causing the linear emission feature we see at $K$. Higher resolution observations may also reveal additional structure, in the linear feature similar to that observed in the spiral arms of the nearby galaxy M51 (i.e. emission and extinction might dominate in different locations). Higher resolution CO data would provide a more accurate estimate of the gas inflow rate. Deeper optical images or a large scale HI map might reveal evidence (such as tidal tails or shells) of the merger that we suggest is responsible for the present appearance of the galaxy. Study of the velocity field in H$\alpha$ could confirm our interpretation of the outer pair of linear features observed in H$\alpha$ as a response to a perturbation at a slower pattern speed than the inner pair.

We hope that future studies of these shock like features observed in barred galaxies are coupled with dynamical models. These studies would provide a more accurate measurement of the actual inflow rates, as well as a way to investigate the response and dissipation of the interstellar medium in these shocks.

We acknowledge helpful discussions and correspondence with L. Athanassoula, L. Kuchinski, A. Gould, J. Turner, R. Hurt, R. González and C. Mihos. We thank R. Bertram for help with the observations in Flagstaff. J. A. F. thanks the staff of the Las Campanas Observatory in Chile for their assistance and Leonard Searle for providing access to the research facilities of the Carnegie Observatories. The OSU galaxy survey is being supported in part by NSF grant AST-9217716. OSIRIS was built with substantial aid from NSF grants AST-9016112 and AST-9218449. IFPS instrument development was supported by NSF Grants AST-8822009 and AST-9112879. The Las Campanas Observatory IR camera was built with partial funding from NSF grant AST-9008937 to S. E. Persson. J.A.F's research is supported in part by NSF grant AST-9218281. A.C.Q. acknowledges the support of a Columbus fellowship. This work used the FITSIO library. We thank the anonymous referee for helpful comments and suggestions which improved the paper.

TABLE 1

Colors [1]

| color | outer disk | out. ridge | lens | nucleus | lin. feature[2] |
|---|---|---|---|---|---|
| $B - J$ | 2.75 | 2.8 | 3.3 | 4.05 | |
| $V - J$ | 2.1 | 2.15 | 2.4 | 3.00 | |
| $R - J$ | 1.6 | 1.7 | 1.9 | 2.35 | |
| $J - H$ | 0.5 | 0.6 | 0.7 | 0.77 | 0.73 |
| $J - K$ | 0.7 | 0.8 | 0.9 | 1.28 | 0.98 |

[1] Estimated errors are typically less than 0.1 mag and are primarily determined by variation of colors in the features, rather than the systematics of the observations. See §3.1 and 3.3 for discussion.

[2] The optical colors are omitted for the linear feature because of the large color variations in this feature.

## REFERENCES

Aaronson, M. 1977, Ph.D. Thesis, Harvard University
Athanassoula, E. 1992, MNRAS, 259, 345
Barnaby, D., & Thronson, M. A. 1992, AJ, 103, 41
Barnes, J. E., & Hernquist, L. E. 1991, ApJ, 370, L65
Beckman, J. E., & Cepa, J. 1990, A&A, 229, 37
Blackman, C. P. 1983, MNRAS, 202, 379
Bloemen, J. B. G. M., Strong, A. W., Mayer-Hasselwander, H. A., Blitz, L., Cohen, R. S., Dame, T. M., Grabelsky, D. A., Thaddeus, P., Hermsen, W., & Lebrun, F. 1986, A&A, 154, 25
Burbidge, E. M., Burbidge, G. R., & Prendergast, K. H. 1960, ApJ, 132, 654
DePoy, D. L., Atwood, B., Byard, P., Frogel, J., & O'Brien, T. 1993, In SPIE Vol 1946, Infrared Detectors and Instrumentation, p. 667
Devereux, N.A. 1989, ApJ, 346, 126
Dressel, L. L., & Condon, J. J. 1976, ApJS, 31, 187
Duval, M. F., & Monnet, G. 1985, A&AS, 61, 141
Elias, J. H., Frogel, J. A., Matthews, K., & Neugebauer, G. 1982, AJ, 87, 1029
Elmegreen, B. G.,& Elmegreen, D. M. 1985, ApJ, 288, 438
Fillipenko, A. V. & Sargent, W. L. W. 1985, ApJS, 57, 503
Frogel, J. A. 1988, ARA&A, 26, 51
Grøsbol, P. J. 1985, A&AS, 60, 261
Kenney, J. D. P. & Lord, S. D. 1991, ApJ, 381, 118
Kenney, J. D. P., Wilson, C. D., Scoville, N. Z., Devereux, N. A., & Young, J. S. 1992, ApJ, 395, L79
Kennicutt, R.C., Jr., & Kent, S.M. 1983, AJ, 88, 1094
Hernquist, L. 1989, Nature, 340, 687
Keel, W. C. 1983, ApJ, 269, 466
Krabbe, A., Sternberg, A., & Genzel, R. 1994, ApJ, 425, 72
Landolt, A. U. 1992, AJ, 104, 340
Lester, D. F., Carr, J. S., Joy, M., & Gaffney, N. 1990, ApJ, 352, 544
Longo, G., & deVaucoulers A. 1983, A General Catalogue of Photoelectric Magnitudes and Colors in the U,B,V System of 3,578 Galaxies Brighter than the 16-th V-magnitude (1936-1982), published by the Department of Astronomy, The University of Texas at Austin
Mihos, J. C., & Hernquist, L. 1994a, ApJ, 425, L13
Mihos, J. C., & Hernquist, L. 1994b, ApJL, 321, L9
Noguchi, M. 1987, MNRAS, 228, 635
Noguchi, M. 1988, A&A, 203, 259
Padin, S. et al. 1991, PASP, 103, 461
Persson, S. E., West, W. C., Carr, D., Sivaramakrishnan, A. & Murphy, D. 1992, PASP, 104, 204
Pogge, R. W. 1992, in Astronomical CCD Observing and Reduction Techniques, S.B. Howell, ed., ASP Conference Series, 23, 195
Pogge, R. W. 1989, ApJS, 71, 433
Quillen, A. C., Frogel, J. A., & González, R. A. 1994, ApJ in press

Roche, P. F., Aiken, D. K., Smith, C. H., & Ward, M. J. 1991, MNRAS, 248, 606
Rieke, G. H, Lebofsky, M. J., Thompson, R. I., Low, F. J., & Tokunaga, A. T. 1980, ApJ, 238, 24
Rix, H.-W., & Rieke M. J. 1993, ApJ, 418, 123
Sholsman, I., Begelman, M. C., & Frank, J. 1990, Nature, 345, 679
Shlosman, I., & Noguchi, M. 1993, ApJ, 414, 474
Rand, R. J., & Kulkarni, S. 1990, ApJ, 349, L43
Rand, R. J., & Tilanus, R. P. J. 1990, in "The Interstellar Medium in Galaxies", Invited talks at the Second Wyoming Conference held at Grand Teton National Park, Wyoming USA, 1989, ed. H.A. Thronson Jr. and J.M. Schull (Dordrect: Kluwer Academic Publishers)
Telesco, C. M., Dressel, L. L., Wolstencroft, R. D. 1993, ApJ, 414, 120
Tully, R. B. 1989, *Nearby Galaxies Catalog*, (Cambridge: Cambridge University Press)
Turner, J. L., Hurt, R. L., & Hudson, D. Y. 1993, ApJ, 413, L19
van Albada, G. D., & Roberts, Jr., W. W. 1981, ApJ, 246, 740
Wainscoat, R. J., Freeman, K. C, & Hyland, A. R. 1989, ApJ, 337, 163
Willner, S. P., Elvis, M., Fabbiano, G., Lawrence, A., & Ward, M. J. 1985, ApJ, 299, 443
Worthey G., 1994, ApJS, in press

## FIGURE CAPTIONS

**Figure 1.** Images at $B$, $V$, $R$, $J$, $H$, and $K$ displayed in a linear scale. North is approximately up and East is approximately to the left. Up is at a position angle (from North) of $+3.6°$. Note that in the optical bands both the effects of dust absorption and young blue stars can be seen. The infrared images have much less small scale structure. The linear streaks in the optical images are from a bright star outside the field of view that bled across the CCD. The scale in arcseconds is shown on the lower left.

**Figure 2.** Images at $B$, $V$, $R$, $J$, $H$, and $K$ in magnitudes. Contours are 0.5 magnitudes per square arcsec apart with maximum contour at 23.5, 23.0, 22.0, 20.5, 20.0, 19.5 magnitudes per square arcsec in $B$,$V$,$R$,$J$,$H$, and $K$ respectively. The scale in arcseconds is shown on the lower left.

**Figure 3.** Color maps, H$\alpha$ and CO(1-0) emission. Black represents redenning in the color maps. The $J-K$, $J-H$, $H-K$, and $R-J$ color maps are displayed from 0.65 to 1.15, 0.4 to 0.9, 0.0 to 0.4 and 1.0 to 2.5 mag respectively. The H$\alpha$ image is displayed as the log of the surface brightness over a range of -16.7 to -14.7 with the H$\alpha$ surface brightness in ergs cm$^{-2}$ s$^{-1}$ arcsec$^{-2}$. The CO image is displayed as the log of the surface brightness over a range of -1.5 to 0.5 with the CO(1-0) surface brightness in Jy km s$^{-1}$ arcsec$^{-2}$. The linear feature running along the bar is easy to see in the $J-K$ and $H-K$ color maps but difficult to see in the $J-H$ color map implying that this feature is very red. This feature is the site of both H$\alpha$ and CO(1-0) emission.

**Figure 4.** Additional color maps to show features on the sides of the bar. The $B-V$, $R-J$, $J-K$, and $B-J$ color maps are displayed from 0.4 to 1.2, 1.0 to 2.5, 0.65 to 1.35 and 2.5 to 4.2 mag respectively. The H$\alpha$ image is displayed as in Figure 3. The lower right hand corner shows a crude representation of the peaks of the H$\alpha$ emission. The outer pair of linear features is coincident with blue colors in the color maps. We refer to the region bounded by these outer linear features as the "lens shaped region".

**Figure 5.** a) H$\alpha$ gray scale and CO(1-0) contour overlay. CO(1-0) contours are at 0.05,0.1,0.2, 0.5,1.0,2.0 Jy km s$^{-1}$ arcsec$^{-2}$. The CO contour map covers a $30'' \times 115''$ region (bordered by the dotted lines) containing the nucleus and bar.

b) H$\alpha$ gray scale and $K$ contour overlay. The lowest contour is 19.2 mag arcsec$^{-2}$ with contours increasing by a ratio of 1.5 in surface brightness.

**Figure 6.** East-West slices across the bar. The dotted line is the surface brightness in the optical $R$ band and the thin solid lines is the surface brightness in the infrared $K$ band. The thick solid line is the H$\alpha$ emission and the dashed line is the CO(1-0) emission. All curves have been normalized so that their peak is 1. Each plot consists of an average across a slice $4''.5$ wide extending in the East-West direction. Each slice is centered at a position directly north or south of the nucleus. Offsets of the center of the slice are given in arcsecs on the right hand side of each plot. The horizontal axis is an offset in arcsecs (in the east west direction) from the center of each slice. Note that there is an offset between the peaks of the H$\alpha$ and CO(1-0 )emission and the peak of the bar as seen in the infrared images.

**Figure 7.** Contours of the gravitational potential derived with vertical scale length $h = 4'' = 640$pc overlayed on a gray scale of the $K$ surface brightness. Contours are $10^4$ (km/s)$^2$ apart with the central contour at $-22 \times 10^4$ (km/s)$^2$. Units of the potential are multiplied by factors given in the text. We haved corrected for the inclination of the galaxy assuming an inclination of 45° and a position angle of 39° (Grøsbol 1985).

**Figure 8.** a) Circular rotation speed, $v_c$, derived from the azimuthal average of the gravitational potential. b) Angular rotation rate $\Omega = v_c/r$ is plotted as the solid line. The dotted line is $\Omega - \kappa/2$, where $\kappa$ is the epicylic frequency. For a bar angular rotation rate $\Omega_b = 0.1\mathrm{Myr}^{-1}$ (shown as the horizontal line), the corotation radius is at $r \sim 45''$.

**Figure 9.** Predicted inflow speed $dr/dt$ as a function of distance from the nucleus. The solid line shows the speed estimated from the location of the CO(1-0) emission, the dotted line the speed estimated from the H$\alpha$ emission.